\documentclass[showpacs,preprintnumbers,amsmath,aps,amssymb]{revtex4}
\usepackage{graphicx}
\usepackage{dcolumn}
\usepackage{bm}
\begin{document}

\baselineskip 18pt

\title{Classical Bianchi type I  cosmology  in  K-essence theory}
\author{J. Socorro$^{1}$ }
\email{socorro@fisica.ugto.mx}
\author{Luis O. Pimentel$^2$}
\email{lopr@xanum.uam.mx}
\author{Abraham Espinoza-Garc\'ia$^1$ }
\email{abraham@fisica.ugto.mx}
\affiliation{$^1$Departamento de F\'{\i}sica de la  DCeI de la  Universidad de
Guanajuato-Campus Le\'on\\
A.P. E-143, C.P. 37150,  Guanajuato, M\'exico\\
$^2$Departamento de F\'{\i}sica de la  Universidad Aut\'onoma Metropolitana\\
Apartado Postal 55-534, 09340, M\'exico, D.F.}
\begin{abstract}
We use one of the simplest forms of the K-essence theory and we
apply it to the classical anisotropic Bianchi type I cosmological
model, with a barotropic perfect fluid ($p= \gamma\rho$)  modeling
the usual matter content and with cosmological constant $\Lambda$,
corresponding at case $\rm V(\phi)=2\Lambda$ in the K-essence
lagrangian density. The classical solutions for any $\gamma\not=1$
and $\Lambda=0$ are found in closed form, using a time
transformation. We also present the solution when $\Lambda \not=0$
including particular values in the barotropic parameter. We present
the possible isotropization of the cosmological model Bianchi I
using the ratio between the anisotropic parameters and the volume of
the universe and show that this tend to a constant or to zero for
different cases. Also we include a qualitative analysis of the
analog of the Friedmann equation when it is written as an equation
for the volume that is equivalent to the equation of motion of a
particle under a potential and we conclude the same about the
isotropization of this anisotropic models for the stiff and
radiation eras of the universe, in the field $\phi$.

Keywords:  K-essence theory; anisotropic model; classical solution; \\

  \end{abstract}

\pacs{02.30.Jr; 04.60.Kz; 12.60.Jv; 98.80.Qc.}
  \maketitle                            

\section{Introduction}
 In recent times, some attempts to unify the description of dark
matter, dark energy and inflation, by means of a scalar field with
non standard kinetic term have been conducted \cite{s-b,armendariz,
bose, jorge,jorge1}.  The K-essence theory   is based on the idea of
a dynamical attractor solution which causes it to act as a
cosmological constant only at the onset of matter domination.
Consequently, K-essence overtakes the matter density and induces
cosmic acceleration at about the present epoch.  Usually K-essence
models are restricted to the lagrangian density of the form
\cite{roland,chiba,bose,arroja,tejeiro}
\begin{equation}
\rm S=\int d^4x \, \sqrt{-g}\,\left[ f(\phi) \, {\cal
G}(X)-V(\phi)\right], \label{kk}
\end{equation}
where the canonical kinetic energy is given by $\rm {\cal
G}(X)=X=-\frac{1}{2}\nabla_\mu \phi \nabla^\mu \phi$. K-essence was
originally proposed as a model for inflation, and then as a model
for dark energy, along with explorations of unifying dark energy and
dark matter \cite{roland,bilic,bento}. Another motivations to
consider this type of lagrangian originates from string theory
\cite{string}. For more details for K-essence applied to dark energy
you can  see \cite{copeland} and references therein.

In this framework, gravitational and matter variables have been
reduced to a finite number of degrees of freedom. For homogenous
cosmological models the metric depends only on time and gives a
model with a finite dimensional configuration space, called
minisuperspace. In this work, we use this formulation to obtain
classical solutions to the anisotropic Bianchi type I  cosmological
model with a perfect fluid. This class of models were considered
initially in this formalism by Chimento and Forte \cite{chimento}.
The first step is to write the  theory for the Bianchi type I model
in the usual manner, that is, we calculate the corresponding
energy-momentum tensor to the scalar field and give the equivalent
Lagrangian density. Next, by means of a Legendre transformation, we
proceed to obtain the canonical Lagrangian ${\cal L}_{can}$, from
which the classical Hamiltonian ${\cal H}$ can be found.

One of the  simplest  K-essence models, without self interaction has
the following lagrangian density
\begin{equation}
\rm {\cal L}_{geo}=\left( \frac{R}{2}+ f(\phi) {\cal
G}(X)\right), \label{lagrangian}
\end{equation}
R is the scalar curvature, $\Lambda$ is the cosmological constant,
and $f(\phi)$ is an arbitrary function of the scalar field.

From the Lagrangian (\ref{lagrangian}) we can build the complete action
\begin{equation}
\rm I=\int_{\Sigma} \sqrt{-g}({\cal L}_{geo}+ {\cal L}_\Lambda+
{\cal L}_{mat})d^4x,\label{action}
\end{equation}
where ${\cal L}_{mat}$ is the matter Lagrangian, ${\cal L}_\Lambda$
is the cosmological constant lagrangian, and g is the determinant of
the metric tensor. The field equations for this theory are
\begin{subequations}
\begin{eqnarray}
\rm G_{\alpha \beta}+ \Lambda g_{\alpha \beta}+ f(\phi) \left[{\cal
G}_X \phi_{,\alpha}\phi_{,\beta}
 +  {\cal G} g_{\alpha \beta}  \right]
&=& -\rm T_{\alpha \beta}, \label{efe} \\
\rm  f(\phi)\left[{\cal G}_X\phi^{,\beta}_{\, ;\beta} + {\cal
G}_{XX}X_{;\beta}\phi^{,\beta} \right] + \frac{df}{d\phi}\left[{\cal
G} - 2X{\cal G}_X\right]&=&0, \label{fe-k}
\end{eqnarray}
\end{subequations}
where we work in units with  $8\pi G=1$  and, as usual, the semicolon
means a covariant derivative and a subscripted X denotes
differentiation with respect to X.

The same set of equations(\ref{efe},\ref{fe-k}) is obtained if we
consider the scalar field $X(\phi)$ as part of the matter content,
i.e. say $\rm {\cal L}_{X,\phi}=\rm f(\phi) {\cal G}(X)$ with the
corresponding energy-momentum tensor
\begin{equation}
\rm {\cal T}_{\alpha \beta}= f(\phi)\left[{\cal G}_X
\phi_{,\alpha}\phi_{,\beta}
 +  {\cal G}(X)g_{\alpha \beta}  \right]. \label{ener-mom}
 \end{equation}

Considering the energy-momentum tensor of a barotropic perfect
fluid,
\begin{equation}
\rm T_{\alpha\beta}=(\rho +P)u_\alpha u_\beta + P g_{\alpha \beta},
\label{tensor-matter}
\end{equation}
with $\rm u_\alpha$ the four-velocity, which satisfy the relation
$\rm u_\mu u^\mu=-1$, $\rho$ the energy density and P the pressure
of the fluid. For simplicity we consider a comoving perfect fluid.
The pressure, the energy density and the four-velocity corresponding
to the energy-momentum tensor of the field X, become
\begin{equation}
\rm P(X)= f(\phi) {\cal G}, \qquad \rho(X)= f(\phi)\left[2X{\cal
G}_X-{\cal G} \right], \qquad u_\mu=\frac{\nabla_u \phi}{\sqrt{2X}},
\label{pX}
\end{equation}
thus, the barotropic parameter is
\begin{equation}
\rm \omega_X=\frac{{\cal G}}{2X{\cal G}_X-{\cal G}}.
\end{equation}
and we notice that the case of a  constant barotropic index $\omega_X$, (with the exception $\omega_X=0$) can be obtained by the
 ${\cal G}$ function
\begin{equation} \rm
{\cal G}=X^{\frac{1+\omega_X}{2\omega_X}}.
\end{equation}
We have the following states in the evolution of our universe in
this formalism,
\begin{equation}
\left\{ \begin{tabular}{ll} stiff matter :& $\rm \omega_X=1$, $\to$
$\rm {\cal G}(X)=X$. \\
Radiation: &$\rm \omega_X=\frac{1}{3}$, $\to$  $\rm {\cal
G}(X)=X^2$.\\
inflation like: & $\rm \omega_X=-\frac{1}{3}$, $\to$ $\rm {\cal
G}(X)=\frac{1}{X}$.\\
&\\ & $\rm \omega_X=-\frac{2}{3}$, $\to$ $\rm {\cal
G}(X)=\frac{1}{\sqrt[4]{X}}$.
\end{tabular}
\right. \nonumber
\end{equation}
 The mathematical
analysis for the two last cases is very complicated in both regimes,
classical and quantum one. For quantum radiation case, the resulting
Wheeler-DeWitt equation appears as fractionary differential
equation, and the results will be reported elsewhere. In reference
\cite{jorge1}, the authors present the analysis to radiation era
using dynamical systems obtaining bouncing solutions.
\subsection{Anisotropic cosmological  Bianchi Class A models, $\rm f(\phi)$=constant}
Considering the cosmological anisotropic Bianchi Class A models with
metric (\ref{met}), the equation (\ref{fe-k}) in term of X, becomes
(here and all where appear the $\prime$ means,
$\prime=\frac{d}{d\tau}=\frac{d}{Ndt}$, with t the usual cosmic
time)

\begin{equation}
\rm \left[{\cal G}_X  + 2 X {\cal G}_{XX} \right]X^\prime +
6\Omega^\prime X{\cal G}_X =0,
 \label{bianchi-k-model}
\end{equation}
and its corresponding solution
\begin{equation}
\rm X{\cal G}_X^2=\eta e^{-6\Omega}.\label{bianchi-k-solution}
\end{equation}
with $\eta$ a constant.

Note that equation
(\ref{bianchi-k-solution}) give  us
the possible solutions $\rm X(A)$, as a function
of the scale factor and therefore the behavior of all
physical properties of the k-essence (like $\rho$, P) ,
are completely determined by the function X and
do not depend on the evolution of the other types of energy density.
The only dependence of the k-essence component on other components
enters through $\rm A(\tau)=e^{\Omega(\tau)}$ in Bianchi Class A
cosmological models.

In the following we present the analysis when  $\rm f(\phi)$ is a
constant and  generic function of the field $\phi$ and assuming a
Bianchi type I metric, which is the anisotropic generalization of
flat FRW cosmological model, and we present the solution in
quadrature form.

\subsubsection{quintessence like case: ${\cal G}=X^2$ and $\rm f(\phi)=constant$}
Using the equation
\begin{equation}
\rm X{\cal G}_X^2=\eta e^{-6\Omega}.
\end{equation}
for the energy kinetic we have the form
\begin{equation} \rm
X=\sqrt[3]{\frac{\eta}{4}} e^{-2\Omega}, \label{ra} \end{equation}
then the field $\phi$ have the solution $$\rm \Delta \phi=\sqrt{2}
\sqrt[6]{\frac{\eta}{4}} \int e^{-\Omega} d\tau.$$

\subsubsection{quintessence like case: ${\cal G}=X$ and $\rm f(\phi)\not=constant$}

The field equations for this particular case are
\begin{subequations}
\begin{eqnarray}
\rm G_{\alpha \beta}+ \Lambda g_{\alpha \beta}+f(\phi) \left(
\phi_{,\alpha}\phi_{,\beta}
 - \frac{1}{2} g_{\alpha \beta} \phi_{,\gamma} \phi^{,\gamma} \right)
&=& \rm - T_{\alpha \beta}, \label{efe-model} \\
\rm 2f(\phi) \phi^{,\alpha}_{\,\,;\alpha} +
\frac{df}{d\phi}\phi_{,\gamma} \phi^{,\gamma}&=&0, \label{fe}
\end{eqnarray}
\end{subequations}
and the  energy-momentum tensor (\ref{ener-mom}) has the following
form,
\begin{equation}
\rm {\cal T}_{\alpha \beta}=f(\phi) \left(
\phi_{,\alpha}\phi_{,\beta}
 - \frac{1}{2} g_{\alpha \beta} \phi_{,\gamma} \phi^{,\gamma} \right). \label{ener-momen}
\end{equation}
In this new line of reasoning, the action (\ref{action}) can be
rewritten as a geometrical part and matter content (usual matter
plus a term that corresponds to the exotic scalar field component of
the K-essence theory).

The equation of motion for the field $\phi$ (\ref{fe}) has the
following property, using the metric of the Bianchi type I model
(however, this is satisfied by all cosmological Bianchi Class A
models),
\begin{equation}
\rm 3\Omega^\prime \phi^\prime f + \phi^{\prime\prime} f +
\frac{1}{2} \frac{df}{d\phi} \phi^{\prime 2}=0,
\end{equation}
which can be integrated at once with the following result,
\begin{equation}
\rm \frac{1}{2}f(\phi) \,\phi^{\prime 2}= \eta e^{-6\Omega}, \qquad
\rightarrow \qquad \int \sqrt{f(\phi)} d\phi=\sqrt{2\eta}\int
e^{-3\Omega(\tau)}d\tau. \label{property}
\end{equation}
here $\eta$  is an integration constant and has the same sign as $f(\phi)$.
 Considering the particular form of $\rm f(\phi)=\omega \phi^m$
or $\rm f(\phi)=\omega e^{m\phi}$ with m and $\omega$ constants, the
classical solutions for the field $\phi$ in quadrature are
\begin{equation}
\rm \phi(\tau)=\left\{
\begin{tabular}{ll}
$\rm \left[ (m+2)\sqrt{\frac{\eta}{2\omega}}\int
e^{-3\Omega}d\tau\right]^{\frac{2}{m+2}}$, & $f(\phi)=\omega
\phi^m, \quad m\not= -2$\\
$\rm Exp\left\{\sqrt{\frac{2\eta}{\omega}}\int
e^{-3\Omega}d\tau\right\}$, & $f(\phi)=\omega
\phi^{-2}, \quad m= -2$\\
$\rm \frac{2}{m}\,Ln\left[m\,\sqrt{\frac{\eta}{2\omega}} \int
e^{-3\Omega}d\tau\right]$, & $f(\phi)=\omega e^{m\phi}, \quad
m\not=0$\\
$\rm \sqrt{2 \eta} \int e^{-3\Omega}d\tau$& $f(\phi)=\omega, \quad
m=0$
\end{tabular}
\right. \label{general-solution-phi}
\end{equation}
in the particular gauge $\rm N=24 e^{3\Omega}$,
(\ref{general-solution-phi}) simplifies to (remember that $\rm
d\tau=N(t)dt$)
\begin{equation}
\rm \phi(t)=\left\{
\begin{tabular}{ll}
$\rm \left[
24(m+2)\sqrt{\frac{\eta}{2\omega}}\,t\right]^{\frac{2}{m+2}}$, &
$f(\phi)=\omega
\phi^m, \quad m\not= -2$\\
$\rm Exp\left\{24\sqrt{\frac{2\eta}{\omega}}\,t\right\}$, &
$f(\phi)=\omega
\phi^{-2}, \quad m= -2$\\
$\rm \frac{2}{m}\,Ln\left[24m\sqrt{\frac{ \eta}{2\omega}}
\,t\right]$, & $f(\phi)=\omega e^{m\phi}, \quad m\not=0$\\
$24\sqrt{2\eta}\,\, t,$ & $f(\phi)=\omega, \quad m=0$
\end{tabular}
\right. \label{gauge-general-solution-phi}
\end{equation}

In our particular case, it
is evident that the contribution of the scalar field is equivalent
to a stiff fluid with a barotropic equation of state $\gamma=1$.
  This is an instance of the results of the
analysis of the energy momentum tensor of a scalar field
(\ref{ener-momen}) by Madsen \cite{marden} for General Relativity
with scalar matter and by Pimentel \cite{pimentel} for the general
scalar tensor theory. In both works a free scalar field is
equivalent to a stiff matter fluid.  In this way, we write the
action (\ref{action}) in the usual form
\begin{equation}
\rm I=\int_{\Sigma} \sqrt{-g}\left( \frac{R}{2}+{\cal L}_\Lambda
+{\cal L}_{mat}+{\cal L}_\phi \right)d^4x,\label{action1}
\end{equation}
and consequently, the classical equivalence between the two theories.
We can infer that this correspondence is also satisfied in the quantum
regime, so we can use this structure for the quantization program, where the
ADM formalism is well known for different classes of matter \cite{ryan1}.

This work is arranged as follows.  In section II we construct the
lagrangian and hamiltonian densities for the anisotropic Bianchi
type I cosmological model. In section III we present some ideas in
as the anisotropic cosmological model can obtain its isotropization
via the mean volume function and next we obtain the classical exact
solution for all values in the gamma parameter. Finally, section IV
is devoted to some final remarks.
\section{Hamiltonian for the Bianchi type I cosmological model}
Let us recall here the canonical formulation in the ADM formalism of the
diagonal Bianchi Class A models. The metric has the form
\begin{equation}
\rm ds^2= -(Ndt)^2 + e^{2\Omega(t)}\, (e^{2\beta(t)})_{ij}\,
\omega^i \, \omega^j=-d\tau^2 + e^{2\Omega(t)}\,
(e^{2\beta(t)})_{ij}\, \omega^i \, \omega^j, \label {met}
\end{equation}
where $\Omega(t)$ is a scalar, N the lapse function and $\rm
\beta_{ij}(t)$ a 3x3 diagonal matrix, $\rm \beta_{ij}= diag(\beta_++
\sqrt{3} \beta_-,\beta_+- \sqrt{3} \beta_-, -2\beta_+)$, $\rm
\omega^i$ are one-forms that  characterize  each cosmological
Bianchi type model and obey $\rm d\omega^i= \frac{}{2} C^i_{jk}
\omega^j \wedge \omega^k,$ $\rm C^i_{jk}$ the structure constants of
the corresponding invariance group. For the Bianchi type I model we
have
$$\rm \omega^1=dx^1\qquad \omega^2=dx^2, \qquad  \omega^3=dx^3$$
 The total Lagrangian density then for this metric becomes
\begin{equation}
  \rm {\cal L}_{I} =\rm  e^{3\Omega} \left[6 \frac{\dot \Omega^2 }{N} - 6 \frac{\dot \beta_+^2}{N}- 6 \frac{\dot \beta_-^2}{N}
  + \frac{f(\phi)}{2N} \dot \phi^2 + 2 N  \rho + 2N \Lambda  \right],
\label{i-lagrangian}
\end{equation}
using the standard definition  of the  momenta, $\rm
\Pi_{q^\mu}=\frac{\partial{\cal L}}{\partial{\dot q^\mu}}$, where
$\rm q^{\mu}=(\Omega, \beta_+,\beta_-, \phi)$, we obtain
\begin{eqnarray*}
\rm \Pi_\Omega&=&\rm \frac{12}{N}e^{3\Omega}\dot \Omega, \quad \rightarrow \quad \dot \Omega=\frac{N}{12}e^{-3\Omega}\Pi_\Omega \\
\rm \Pi_+&=& \rm -\frac{12}{N}e^{3\Omega}\dot \beta_+, \quad \rightarrow \quad \dot \beta_+=-\frac{N}{12}e^{-3\Omega}\Pi_+\\
\rm \Pi_-&=& \rm -\frac{12}{N}e^{3\Omega}\dot \beta_-, \quad \rightarrow \quad \dot \beta_-=-\frac{N}{12}e^{-3\Omega}\Pi_+\\
\rm \Pi_\phi&=& \rm \frac{f}{N}e^{3\Omega}\dot \phi, \quad \rightarrow \quad \dot \phi=\frac{N}{f}e^{-3\Omega}\Pi_\phi\\
\end{eqnarray*}
and introducing them into the Lagrangian density, we obtain the canonical
Lagrangian as $\rm {\cal L}_{canonical}=\Pi_{q^\mu} \dot q^\mu -N
{\cal H}$,
\begin{eqnarray}
\rm {\cal L}_{canonical}&=&\rm \Pi_{q^\mu} \dot q^\mu
-\frac{N}{24}e^{-3\Omega} \left\{ \Pi_\Omega^2 +
\frac{12}{f(\phi)}\Pi_\phi^2 -\Pi_+^2 -\Pi_-^2 -48\mu_\gamma
e^{-3(\gamma-1)\Omega}-48\Lambda e^{ 6\Omega}\right\},
\end{eqnarray}
where we have used  energy-momentum conservation the law for a
perfect fluid, $\rm T^{\mu \nu}_{;\nu}=0, \to \rho=\mu_\gamma
e^{-3(\gamma+1)\Omega}$, we assumed an equation of state $\rm
p=\gamma \rho$, so, the corresponding Hamiltonian density is

\begin{equation}
{\cal H}_\perp= \rm \frac{e^{-3\Omega}}{24}\left(-\Pi^2_\Omega
-\frac{12}{f(\phi)}\Pi_\phi^2+ \Pi^2_+ +\Pi^2_- +
b_{\gamma}e^{-3(\gamma-1)\Omega}+48\Lambda e^{ 6\Omega}\right ),
\label {ham}
\end{equation}
with $\rm b_\gamma=48 \mu_\gamma$.


\subsection{Classical equations}
The corresponding Einstein field equations (\ref{efe-model}) and
(\ref{fe}) for the anisotropic cosmological model bianchi type I are
the following (remember that the prime $'$ is the derivative over
the time $\rm d\tau=Ndt$),
\begin{eqnarray}
&& 3\Omega^{\prime 2}-3\beta_+^{\prime 2}-3\beta_-^{\prime 2}- \frac{f}{4}\phi^{\prime 2}-\rho -\Lambda=0 \nonumber\\
&&\rm 2\Omega^{\prime\prime}+3\Omega^{\prime 2}-3\Omega^{\prime}
\beta_+^{\prime}-3\sqrt{3}\Omega^{\prime} \beta_-^{\prime}
-\beta_+^{\prime\prime} +3\beta_+^{\prime 2}
-\sqrt{3}\beta_-^{\prime\prime}+3\beta_-^{\prime 2}  + \frac{f}{4}\phi^{\prime 2}+p -\Lambda =0, \label{ecuacion-einstein}\\
&&\rm 2\Omega^{\prime \prime}+3\Omega^{\prime
2}-3\Omega^{\prime}\beta_+^{\prime}+3\sqrt{3}\Omega^{\prime}\beta_-^{\prime}
-\beta_+^{\prime \prime}+3\beta_+^{\prime 2}
 +\sqrt{3}\beta_-^{\prime \prime}
+3\beta_-^{\prime 2}  + \frac{f}{4}\phi^{\prime 2}
+ p -\Lambda  =0, \nonumber\\
&&\rm 2\Omega^{\prime \prime}+3\Omega^{\prime
2}+6\Omega^{\prime}\beta_+^{\prime} +2\beta_+^{\prime
\prime}+3\beta_+^{\prime 2} +3\beta_-^{\prime 2}+
\frac{f}{4}\phi^{\prime 2}+ p -\Lambda  =0 \nonumber \\
&& \rm f\left(3\Omega^\prime \phi^\prime  + \phi^{\prime\prime}
\right) + \frac{1}{2} \frac{df}{d\phi} \phi^{\prime 2}=0,
\end{eqnarray}
the solution of this last equation was putted in (\ref{property}),
\begin{equation}
\rm \frac{1}{2}f(\phi) \,\phi^{\prime 2}= \eta e^{-6\Omega}, \qquad
\rightarrow \qquad \int \sqrt{f(\phi)} d\phi=\sqrt{2\eta}\int
e^{-3\Omega(\tau)}d\tau.
\end{equation}
The combination between the second and third equations us give the
solution for the anisotropic function $\beta_-$, also the sum of
third and fourth equations, putting the $\beta_-$ solution, give us
the form of the $\beta_+$ function,
\begin{equation}
\rm \beta_\pm(\tau)=a_\pm \int e^{-3\Omega(\tau)}d\tau, \label{b+-}
\end{equation}
where $\rm a_\pm$ are integration constants. So, the
(\ref{ecuacion-einstein}) are rewritten as

\begin{eqnarray}
\rm 3\Omega^{\prime 2}&=&\rm 3c_1e^{-6\Omega}+\mu_\gamma e^{-3(\gamma+1)\Omega} +\Lambda, \qquad c_1=a_+^2+a_-^2+ \frac{\eta}{6}, \nonumber\\
&&\rm 2\Omega^{\prime\prime}+3\Omega^{\prime 2}+3\beta_+^{\prime 2}
+3\beta_-^{\prime 2}  + \frac{f}{4}\phi^{\prime 2}+ p -\Lambda =0,
\label{ecuacion-modified}
\end{eqnarray}
\subsection{Isotropization}
The current observations of the cosmic background radiation set a
very stringent limit to the anisotropy of the universe \cite{aniso},
therefore it is important to consider the anisotropy of the
solutions. Recalling the Friedmann equation (constraint equation),
\begin{equation}
3\Omega^{\prime 2}-3\beta_+^{\prime 2}-3\beta_-^{\prime 2}-
\frac{f}{4}\phi^{\prime 2}-\rho -\Lambda=0, \label{ome}
\end{equation}
 we can see that
isotropization is achieved when the terms with $\beta_{\pm}^{\prime
2}$ go to zero or are negligible with respect to the other terms in
the differential equation. We find in the literature the criteria
for isotropization, among others, $(\beta_+^{\prime
2}+\beta_-^{\prime 2})/H^2\quad \rightarrow \quad 0$ ,
$(\beta_+^{\prime 2}+\beta_-^{\prime 2})/\rho \quad \rightarrow
\quad  0$ , that are consistent with our above remark. In the
present case the comparison with the density should include the
contribution of the scalar field.  We define an anisotropic density
$\rho_a$, that is proportional to the shear scalar,
\begin{equation}
\rho_a = \beta_+^{\prime 2}+\beta_-^{\prime 2},
\end{equation}
and will compare it with $\rho_\gamma$,$\rho_\phi$, and
$\Omega^{\prime 2}$. From the Hamilton Jacobi analysis we now that
\begin{equation}
\rho_a \sim  e^{-6\Omega}, \qquad \rho_\phi \sim e^{-6\Omega},\qquad
\Omega^{\prime 2} \sim   48 \Lambda+{\kappa_\Omega
}^2e^{-6\Omega}+b_{\gamma}e^{-3(1+\gamma)\Omega}
\end{equation}
and the ratios are
\begin{equation}
\frac{\rho_a }{\rho_\phi}\sim  constant,\qquad \frac{\rho_a
}{\rho_\gamma}\sim      e^{3\Omega(\gamma-1)},\qquad \frac{\rho_a
}{\Omega^{\prime 2      }}\sim \frac{1}{{\kappa_\Omega }^2  +    48
\Lambda e^{6\Omega}+b_{\gamma}e^{3(1-\gamma)\Omega}}.
\end{equation}

Here we see that for an expanding  universe the anisotropic density
is dominated by the fluid density (with the exception of the stiff
fluid) or  by the $\Omega^{\prime 2}$ term and then at late times
the isotropization is obtained if the expansion goes to infinity.
Hence it is necessary to determine when we have an ever expanding
universe. Equation (\ref{ome}) in the new variable $\rm V= e^{3
\Omega}$ is

\begin{equation}
16 V^{\prime 2}  -b_{\gamma } V^{1-\gamma }-48
   \Lambda  V^2 = {\kappa_\Omega }^2, \qquad {\kappa_\Omega }^2= {\kappa_+ }^2+{\kappa_- }^2
+{\kappa_\phi}^2
\end{equation}

 That is equivalent to the equation of motion
in the coordinate $V$ of a particle under the potential $U$ with
energy $E={\kappa_\Omega }^2  $, where
\begin{equation}
U(V) =  -b_{\gamma } V^{1-\gamma }-48
   \Lambda  V^2, \label{poten} .
\end{equation}
We can now have a qualitative idea of the different solutions from
energy diagrams. We assume that $b_\gamma$ is non negative since it
is proportional to the energy density of the fluid. On the other
hand, $\Lambda $ and $ {\kappa_\Omega }^2$ are  real and can take
positive, null or   negative values.
\begin{figure} [h]
\begin {center}
\includegraphics[totalheight=0.25\textheight]{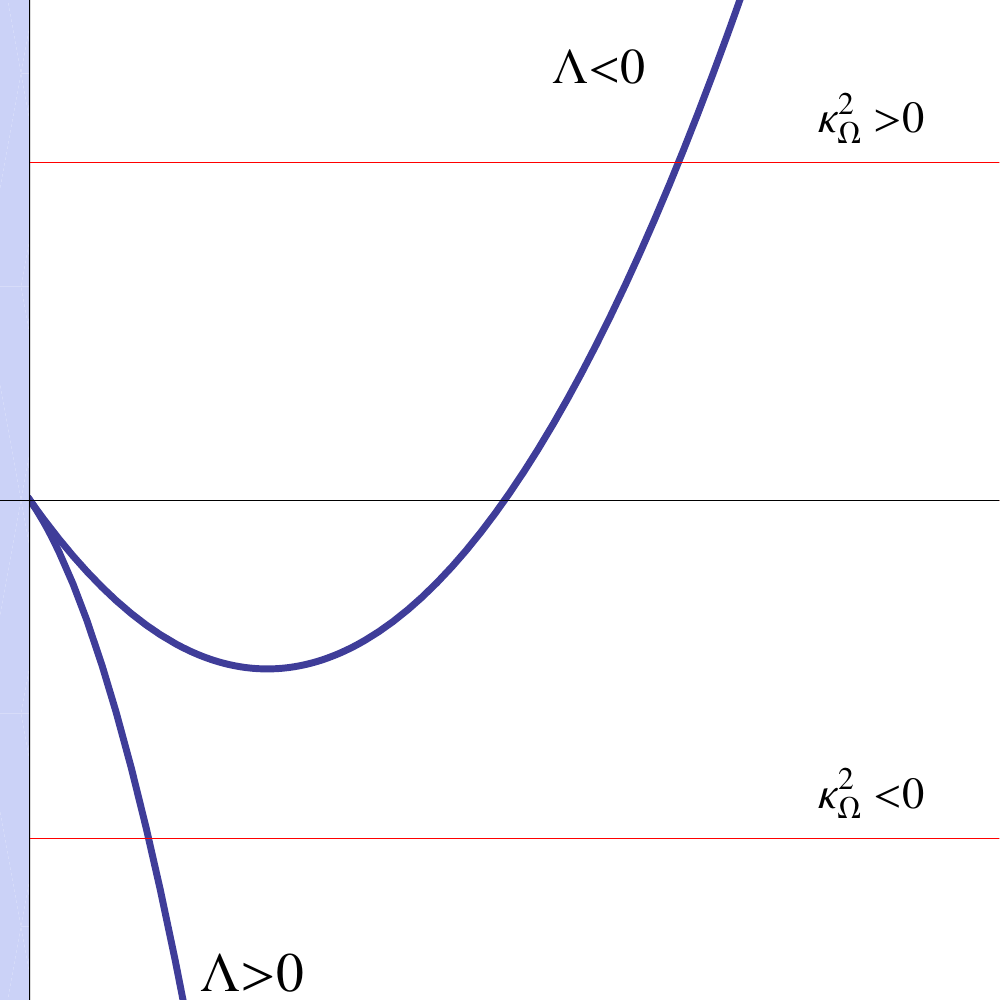}
\caption{Plot of equation (\ref{poten}), considering different
classes of total energy, in the ${\kappa_\Omega }^2 $ parameter.
Also, we include two branches in the cosmological constant. }
\end{center}
\end{figure}

From the figure, that is qualitatively correct for $\gamma \ne 1$,
we see that for negative $\Lambda$ all the expanding solutions will
re-collapse eventually, regardless of the sign of ${\kappa_\Omega
}^2 $.  for positive $\Lambda$ an expanding solution will expand
forever. We also note that, when  ${\kappa_\Omega }^2 <0 $, i.e.
when the {\it ghost} contribution is dominant,  there are
contracting solutions that reach a minimum and then expand forever,
these solutions do not have a big bang  singularity.

\subsubsection{radiation case, ${\cal G}=X^2$}
Reproducing the set of equation for this case, we have that the
Friedmann like  equation is
\begin{equation}\rm
3\Omega^{\prime 2}-3\beta_+^{\prime 2}-3\beta_-^{\prime 2}-
\frac{3}{4}\phi^{\prime 4}-\rho -\Lambda=0, \label{ome-1}
\end{equation}
and making the same analysis that in the previous case, we have
\begin{equation}
\rm 3\Omega^{\prime 2}= 3c_1e^{-6\Omega}+3\phi_1 e^{-4\Omega} +\mu_\gamma e^{-3(\gamma+1)\Omega} +\Lambda, \qquad c_1=a_+^2+a_-^2. \label{ome2}\\
\end{equation}
also the relation between the anisotropic function $\rm \beta_\pm$,
(\ref{b+-}) are satisfied. In the last equation we have used the
equation (\ref{ra}).

Also, we can follow the same structure that for the matter case and
following the Hamilton Jacobi analysis, we now that
\begin{equation}
\rho_a \sim  e^{-6\Omega}, \qquad \rho_\phi \sim e^{-4\Omega},\qquad
\Omega^{\prime 2} \sim   48 \Lambda+\phi_1
e^{-4\Omega}+{\kappa_\Omega
}^2e^{-6\Omega}+b_{\gamma}e^{-3(1+\gamma)\Omega}
\end{equation}
and the corresponding ratios are
\begin{equation}
\frac{\rho_a }{\rho_\phi}\sim e^{-2\Omega}, ,\qquad \frac{\rho_a
}{\rho_\gamma}\sim      e^{3\Omega(\gamma-1)},\qquad \frac{\rho_a
}{\Omega^{\prime 2 }}\sim \frac{1}{{\kappa_\Omega }^2  +    48
\Lambda e^{6\Omega}+\phi_1
e^{2\Omega}+b_{\gamma}e^{3(1-\gamma)\Omega}}.
\end{equation}

 Hence it is
necessary to determine when we have an ever expanding universe.
Equation (\ref{ome2}) in the new variable $\rm V= e^{3 \Omega}$
\begin{equation}\rm
16 V^{\prime 2}  -b_{\gamma } V^{1-\gamma }-48
   \Lambda  V^2 - c V^{2/3}= {E_\Omega}^2, \qquad {E_\Omega }^2= {\kappa_+ }^2+{\kappa_- }^2
\end{equation}

 That is equivalent to the equation of motion
in the coordinate $V$ of a particle under the potential $U$ with
energy $\rm E={E_\Omega }^2  $, where
\begin{equation}\rm
U(V) =  -b_{\gamma } V^{1-\gamma }-48
   \Lambda  V^2 - c V^{2/3}, \label{poten13} .
\end{equation}
for this case, the qualitative analysis is the same that when
 ${\cal G}=X$.

In the following we obtain exact solutions in order to gives the
volume function V to each case to  ${\cal G}=X$.
\subsection{Exact Classical solutions}

In order to find the solutions for the remaining minisuperspace
variables we employ the Einstein-Hamilton-Jacobi equation, which
arises by making the identification $\rm\frac{\partial
S(\Omega,\beta_{\pm},\phi)}{\partial q^{\mu}}=\Pi_{\mu}$ in the
Hamiltonian constraint $\mathcal{H}_{\perp}=0$, which results in
\begin{equation}
\rm\left(\frac{\partial
S}{\partial\Omega}\right)^{2}-\left(\frac{\partial
S}{\partial\beta_{+}}\right)^{2}-\left(\frac{\partial
S}{\partial\beta_{-}}\right)^{2}+\frac{12}{f(\phi)}\left(\frac{\partial
S}{\partial\phi}\right)^{2}-b_{\gamma}e^{3\Omega(1-\gamma)}-48\Lambda
e^{6\Omega}=0\label{hje}
\end{equation}
in order to solve the above equation, we assume a solution of the
form
$S(\Omega,\beta_{\pm},\phi)=S_{1}(\Omega)+S_{2}(\beta_{+})+S_{3}(\beta_{-})+S_{4}(\phi)$
which results in the following set of ordinary differential
equations
\begin{align}
&\rm\left(\frac{dS_{1}}{d\Omega}\right)^{2}-\left(b_{\gamma}e^{-3(\gamma-1)\Omega} +48\Lambda e^{6\Omega}+          {\kappa_\Omega }^2\right)=0\label{hjomega}\\
&\rm\left(\frac{dS_{2}}{d\beta_{+}}\right)^{2}- {\kappa_+ }^2 =0\label{hj+}\\
&\rm\left(\frac{dS_{3}}{d\beta_{-}}\right)^{2}-  {\kappa_- }^2    =0\label{hj-}\\
&\rm\frac{12}{f(\phi)}\left(\frac{dS_{4}}{d\phi}\right)^{2}-   {\kappa_\phi}^2   =0.\label{hjphi}
\end{align}
Here the ${\kappa_i}$ are separation constants satisfying the
relation ${\kappa_\Omega }^2= {\kappa_+ }^2+{\kappa_- }^2
+{\kappa_\phi}^2 $,  ${\kappa_\pm }$ are real,   ${\kappa_\phi}^2 $
should have the same signs as ${f(\phi)}$ and for consistency with
Eq.(\ref{property}) we have $ {\kappa_\phi}^2   = 24\eta$. Recalling
the expressions for the momenta we can obtain solutions for
equations (\ref{hjomega}-\ref{hjphi}) in  quadrature, in particular
\begin{equation} \rm\Delta\tau=12\int\frac{d\Omega}{ \sqrt{48
\Lambda+{\kappa_\Omega }^2e^{-6\Omega}+b_{\gamma}e^{-3(1+\gamma)\Omega}}}.\label{omega-sol}
\end{equation}
\begin{equation}\rm
\Delta\beta _\pm=\mp \frac{\kappa_\pm }{12}\int e^{-3\Omega(\tau)}
d\tau. \label{betas}
\end{equation}

We already know the solution for (\ref{hjphi}). As can be seen from
(\ref{betas}), in order to obtain
solutions for $\beta_{\pm}$ one needs to find a solution for
$\Omega$, which can be obtained from (\ref{omega-sol}).

 The equation (\ref{omega-sol}), does not have a general solution, however, it is possible find solutions for particular values of the
 barotropic parameter  $\gamma$ with $\Lambda \not=0$.

\begin{enumerate}
\item{} {$\Lambda=0$ and $\gamma\not=1$}

The equation (\ref{omega-sol}) can be written as
\begin{equation}
\rm d\tau=12\frac{e^{3\Omega}d\Omega}{
\sqrt{{\kappa_\Omega }^2+b_{\gamma}e^{-3\Omega(\gamma-1)}}}.\label{omega-sol-mod}
\end{equation}
when we consider the time  transformations $\rm d\tau=e^{3\gamma
\Omega} dT$, and the change of variable $\rm u={\kappa_\Omega }^2+b_\gamma
e^{-3(\gamma-1)\Omega}$, this equation has the solution
\begin{equation}
\rm \Omega(T)= Ln\left[\theta_\gamma T^2 + \delta_\gamma T
\right]^{-\frac{1}{3(\gamma-1)}},
\end{equation}
where $\theta_\gamma=\left(\frac{\gamma-1}{8} \right)^2 b_\gamma$
and $\delta_\gamma=-\sqrt{{\kappa_\Omega }^2}\frac{\gamma-1}{4}$. With this, the
time  transformation becomes
$$\rm d\tau= \left[\theta_\gamma T^2 + \delta_\gamma T
\right]^{-\frac{\gamma}{(\gamma-1)}}\, dT.$$ and the closed form is
\cite{andrei},
\begin{equation}
\tau =\frac{(1-\gamma )}{\delta_\gamma } \left[
   \theta_\gamma
   T^2 +\delta_\gamma T\right]^{\frac{1}{1-\gamma }}
   \,
   _2F_1\left(1,-\frac{2}{\gamma -1};\frac{\gamma
   -2}{\gamma -1};-\frac{T
   \theta_\gamma }{\delta_\gamma
   }\right),
\end{equation}
here $_2F_1$ is a hypergeometric function. We also have
\begin{equation}\rm
\rm \int e^{-3\Omega}d\tau=\frac{1}{\delta_\gamma}\,
Ln \left[ \frac{T }{ \theta_\gamma T+  \delta_\gamma}      \right].
\end{equation}
The anisotropy functions and the scalar field are given by
\begin{equation}\rm
\Delta\beta _\pm=\mp \frac{\kappa_\pm }{12  \delta_\gamma}\,
Ln \left[ \frac{T }{ \theta_\gamma T+  \delta_\gamma}      \right].
\end{equation}
\begin{equation}
\rm \phi(T)=\left\{
\begin{tabular}{ll}
$\rm \left[ (m+2)\sqrt{\frac{\eta}{2\omega}}\frac{1}{\delta_\gamma}\,
Ln \left[ \frac{T }{ \theta_\gamma T+  \delta_\gamma}      \right]      \right]^{\frac{2}{m+2}}$, & $f(\phi)=\omega
\phi^m, \quad m\not= -2$\\
$\rm Exp\left\{\sqrt{\frac{2\eta}{\omega}}  \frac{1}{\delta_\gamma}\,
Ln \left[ \frac{T }{ \theta_\gamma T+  \delta_\gamma}      \right]   \right\}$, & $f(\phi)=\omega
\phi^{-2}, \quad m= -2$\\
$\rm \frac{2}{m}\,Ln\left[m\sqrt{\frac{ \eta}{2\omega}}
\frac{1}{\delta_\gamma}\, Ln \left[ \frac{T }{ \theta_\gamma T+
\delta_\gamma}      \right]     \right]$, & $f(\phi)=\omega
e^{m\phi}, \quad
m\not=0$\\
$\rm \sqrt{2 \eta}  \frac{1}{\delta_\gamma}\, Ln \left[ \frac{T }{
\theta_\gamma T+  \delta_\gamma}      \right]     $&
$f(\phi)=\omega, \quad m=0$
\end{tabular}
\right.
\end{equation}

As a concrete example we consider  the particular value $\gamma=0$, then  $\tau=T$ and $\Omega$ becomes
\begin{equation}\rm
\Omega(\tau)=Ln\left[\frac{3}{4} \mu_0 T^2 +
\frac{\sqrt{{\kappa_\Omega }^2}}{4} T \right]^{\frac{1}{3}}, \qquad
\Rightarrow \qquad e^{3\Omega}=\frac{3}{4} \mu_0 T^2 +
\frac{\sqrt{{\kappa_\Omega }^2}}{4} T
\end{equation}
if isotropization is possible, this volume function would not make it quick,
and the integral
$$\rm \int
e^{-3\Omega}d\tau=\frac{4}{\sqrt{{\kappa_\Omega }^2}}\, Ln \left[
\frac{\tau}{\frac{\sqrt{{\kappa_\Omega }^2}}{4}+\frac{3}{4} \mu_0
\tau}\right].$$

So, the classical solutions for the anisotropic function $\beta_\pm$
and $\phi$ field are,
\begin{equation}\rm
\Delta\beta _\pm=\mp \frac{\kappa_\pm }{3\sqrt{{\kappa_\Omega
}^2}}\, Ln \left[ \frac{\tau}{\frac{\sqrt{{\kappa_\Omega
}^2}}{4}+\frac{3}{4} \mu_0 \tau}\right]. \label{betas1}
\end{equation}

\begin{equation}
\rm \phi(\tau)=\left\{
\begin{tabular}{ll}
$\rm \left[ (m+2)\sqrt{\frac{\eta}{2\omega}}Ln \left[
\frac{\tau}{\frac{\sqrt{{\kappa_\Omega }^2}}{4}+\frac{3}{4} \mu_0
\tau}\right]\right]^{\frac{2}{m+2}}$, & $f(\phi)=\omega
\phi^m, \quad m\not= -2$\\
$\rm Exp\left\{\sqrt{\frac{2\eta}{\omega}}Ln \left[
\frac{\tau}{\frac{\sqrt{{\kappa_\Omega }^2}}{4}+\frac{3}{4} \mu_0
\tau}\right]\right\}$, & $f(\phi)=\omega
\phi^{-2}, \quad m= -2$\\
$\rm \frac{2}{m}\,Ln\left[m\sqrt{\frac{ \eta}{2\omega}} Ln \left[
\frac{\tau}{\frac{\sqrt{{\kappa_\Omega }^2}}{4}+\frac{3}{4} \mu_0
\tau}\right]\right]$, & $f(\phi)=\omega e^{m\phi}, \quad m\not=0$\\
$\sqrt{2\eta}\frac{4}{\sqrt{{\kappa_\Omega }^2}}\, Ln \left[
\frac{\tau}{\frac{\sqrt{{\kappa_\Omega }^2}}{4}+\frac{3}{4} \mu_0
\tau}\right]. $& $f(\phi)=\omega, \quad m=0$
\end{tabular}
\right. \label{solution-phi}
\end{equation}

\item{} {$\Lambda =0$ and $\gamma=1$}

In this case equation (\ref{omega-sol}) is
\begin{equation}\rm
\Delta \tau =\int \frac{12}{\sqrt{b_2e^{- 6\Omega} }} d\Omega
\label{int2}
\end{equation}
with $\rm b_2={\kappa_\Omega }^2+48\mu_1$ that we assume positive.
Integrating we obtain
\begin{equation}\rm
e^{ 3 \Omega}= \frac{\sqrt{b_2}\Delta \tau}{4 }.
\end{equation}
if isotropization is possible, this volume function would not make it quick.
For the anisotropic functions we have
\begin{equation}\rm
\Delta\beta _\pm=\mp \frac{\kappa_\pm }{12}\int
e^{-3\Omega(\tau)}d\tau = \mp {\kappa_\pm
}\frac{1}{3\sqrt{b_2}}Ln(\Delta \tau),
\end{equation}
and  the scalar field is given by
\begin{equation}
\rm \phi(\tau)=\left\{
\begin{tabular}{ll}
$\rm \left[
(m+2)\sqrt{\frac{\eta}{2\omega}}\frac{4}{\sqrt{b_2}}Ln(\Delta
\tau)\right]^{\frac{2}{m+2}}$, & $f(\phi)=\omega
\phi^m, \quad m\not= -2$\\
$\rm
Exp\left\{\sqrt{\frac{2\eta}{\omega}}\frac{4}{\sqrt{b_2}}Ln(\Delta
\tau)\right\}$, & $f(\phi)=\omega
\phi^{-2}, \quad m= -2$\\
$\rm \frac{2}{m}\,Ln\left[m\sqrt{\frac{
\eta}{2\omega}}\frac{4}{\sqrt{b_2}}Ln(\Delta \tau) \right]$, &
$f(\phi)=\omega e^{m\phi}, \quad
m\not=0$\\
$\rm \sqrt{2 \eta}\frac{4}{\sqrt{b_2}}Ln(\Delta \tau) $,&
$f(\phi)=\omega, \quad m=0$
\end{tabular}
\right. 
\end{equation}

\item{} {$\Lambda\not=0$ and $\gamma=-1$}

(\ref{omega-sol}) has the form
\begin{equation}\rm
\Delta \tau = \int \frac{12}{\sqrt{{\kappa_\Omega }^2e^{-6\Omega}+b_3}} d\Omega
\label{int1}
\end{equation}
where $\rm b_3=48 \mu_{-1}+48\Lambda$ .
\begin{equation}\rm
\Delta \tau=\frac{4}{\sqrt{b_3}}\,arccsch
\left(\sqrt{\frac{{\kappa_\Omega }^2}{b_3}} e^{-3\Omega}\right)\label{sint1}
\end{equation}
solving for $\Omega$
\begin{equation}\rm
\Omega=\frac{1}{3}Ln\left|\sqrt{\frac{{\kappa_\Omega }^2}{b_3}}\,\sinh\left(
\frac{\sqrt{b_3}}{4}\Delta\tau\right)\right|
\end{equation}
the inverse  volume function is
\begin{equation*}\rm
e^{-3\Omega}=\sqrt{\frac{b_3}{{\kappa_\Omega }^2}}\,csch
\left(\frac{\sqrt{b_3}}{4}\Delta\tau\right)
\end{equation*}
if isotropization is possible, the corresponding volume function would make it quick,
its integral become
\begin{equation}\rm \int
e^{-3\Omega}d\tau=\frac{4}{\sqrt{{\kappa_\Omega }^2}}Ln\left| \tanh
\left(\frac{\sqrt{b_3}}{4}\Delta\tau\right)\right|\label{inv1}
\end{equation}
and as the anisotropic function dependent of this integral, so
\begin{equation}\rm
\Delta \beta_\pm(\tau)=  \pm\frac{\kappa_\pm }{3\sqrt{{\kappa_\Omega }^2}}\,
Ln\left| \tanh \left(\frac{\sqrt{b_3}}{4}\Delta\tau\right)\right|
\end{equation}

Also, the field $\phi$ use this integral (see equation
(\ref{general-solution-phi}), and the corresponding solutions
becomes
\begin{equation}
\rm \phi(\tau)=\left\{
\begin{tabular}{ll}
$\rm \left[ -(m+2)\sqrt{\frac{\eta}{2\omega}}\frac{4}{\sqrt{{\kappa_\Omega }^2}}
Ln\left| \tanh
\left(\frac{\sqrt{b_3}}{4}\Delta\tau\right)\right|\right]^{\frac{2}{m+2}}$,
& $f(\phi)=\omega
\phi^m, \quad m\not= -2$\\
$\rm Exp\left\{-\sqrt{\frac{2\eta}{\omega}}\frac{4}{\sqrt{{\kappa_\Omega }^2}}
Ln\left| \tanh
\left(\frac{\sqrt{b_3}}{4}\Delta\tau\right)\right|\right\}$, &
$f(\phi)=\omega
\phi^{-2}, \quad m= -2$\\
$\rm \frac{2}{m}\,Ln\left[-m\sqrt{\frac{\eta}{2\omega}}
\frac{4}{\sqrt{{\kappa_\Omega }^2}} Ln\left| \tanh
\left(\frac{\sqrt{b_3}}{4}\Delta\tau\right)\right|\right]$, &
$f(\phi)=\omega e^{m\phi}, \quad
m\not=0$\\
$\rm -\sqrt{2 \eta} \frac{4}{\sqrt{{\kappa_\Omega }^2}} Ln\left|
\tanh \left(\frac{\sqrt{b_3}}{4}\Delta\tau\right()\right|$&
$f(\phi)=\omega, \quad m=0$
\end{tabular}
\right. \label{general-solution-phi--1}
\end{equation}
with the condition  $\omega>0$.

Also we can consider the special case, when $\rm b_3=0$ or $\rm
\mu_{-1}=-\Lambda$

\begin{eqnarray}
\rm \Delta \tau&=&\rm \frac{4}{\sqrt{{\kappa_\Omega }^2}}e^{3\Omega}, \nonumber\\
\rm \Omega&=& \rm \frac{1}{3} Ln \left|\frac{\sqrt{{\kappa_\Omega }^2}}{4} \Delta
\tau \right| \nonumber
\end{eqnarray}
 we have the following expression
 \begin{equation} \rm
e^{-3\Omega}= \frac{4}{\sqrt{{\kappa_\Omega }^2}}\frac{1}{\Delta \tau}, \nonumber
\end{equation}
the isotropization is possible for this case, because the
corresponding volume function would make it quick, its integral
become
\begin{equation}
\rm \int e^{-3\Omega}d\tau= \frac{4}{\sqrt{{\kappa_\Omega }^2}} Ln \left|\Delta
\tau \right|.
\end{equation}
so, the anisotropic function $\beta_\pm$ and the $\phi$ function
becomes for this case
\begin{equation}\rm
\Delta \beta_\pm(\tau)=\mp \frac{\kappa_\pm }{3\sqrt{{\kappa_\Omega }^2}}Ln\left|
\Delta \tau\right|.
\end{equation}

\begin{equation}
\rm \phi(\tau)=\left\{
\begin{tabular}{ll}
$\rm \left[
(m+2)\sqrt{\frac{\eta}{2\omega}}\frac{4}{\sqrt{{\kappa_\Omega }^2}}Ln\left| \Delta
\tau\right|\right]^{\frac{2}{m+2}}$, & $f(\phi)=\omega
\phi^m, \quad m\not= -2$\\
$\rm
Exp\left\{\sqrt{\frac{2\eta}{\omega}}\frac{4}{\sqrt{{\kappa_\Omega }^2}}Ln\left|
\Delta \tau\right|\right\}$, & $f(\phi)=\omega
\phi^{-2}, \quad m= -2$\\
$\rm \frac{2}{m}\,Ln\left[m\sqrt{\frac{ \eta}{2\omega}}
\frac{4}{\sqrt{{\kappa_\Omega }^2}}Ln\left| \Delta \tau\right|\right]$, &
$f(\phi)=\omega e^{m\phi}, \quad
m\not=0$\\
$\rm \sqrt{2 \eta} \frac{4}{\sqrt{{\kappa_\Omega }^2}}Ln\left|
\Delta \tau\right|$& $f(\phi)=\omega, \quad m=0$
\end{tabular}
\right. \label{general-solution-phi00}
\end{equation}

\item{} {$\Lambda\not=0$ and $\gamma=0$}

For this case, equation (\ref{omega-sol}) is
\begin{eqnarray}
\rm \Delta \tau&=&\rm \int \frac{12e^{3\Omega} }{\sqrt{{\kappa_\Omega }^2+48\mu_0e^{ 3\Omega } +48\Lambda e^{ 6\Omega}}}d\Omega\\
&=&\rm  \frac{1}{\sqrt{3\Lambda}} Ln\,\left[ \frac{b_0 +96\Lambda
e^{3\Omega}}{48\Lambda} + \frac{1}{2\sqrt{3\Lambda}}\sqrt{{\kappa_\Omega }^2+b_0
e^{3\Omega}+48\Lambda e^{6\Omega}}\right]
\end{eqnarray}
with $b_0=48 \mu_0$, and $\Lambda>0$.

The function $\Omega$ become
\begin{equation}
\rm
\Omega=\frac{1}{3}Ln\,\left[\frac{12\Lambda\left(e^{\sqrt{3\Lambda}\Delta
\tau}-\frac{b_0}{48\Lambda}\right)^2- {\kappa_\Omega }^2}{48 \Lambda
e^{\sqrt{3\Lambda}\Delta \tau}} \right]
\end{equation}
and we have
\begin{eqnarray}
\rm e^{-3\Omega}&=&\rm
\frac{48\Lambda\,e^{\sqrt{3\Lambda}\Delta \tau}}{12\Lambda \left(e^{\sqrt{3\Lambda}\Delta \tau}-\frac{b_0}{48\Lambda} \right)^2-{\kappa_\Omega }^2}\nonumber\\
\rm \int e^{-3\Omega}d\tau&=&\rm -\frac{8}{\sqrt{{\kappa_\Omega }^2}}arctanh\,
\left(2\sqrt{\frac{3\Lambda}{{\kappa_\Omega }^2}}\left(
 -\frac{b_0}{48\Lambda}+e^{\sqrt{3\Lambda}\Delta \tau}\right) \right)\label{int0}
\end{eqnarray}
also, the  isotropization is possible, due that the corresponding
volume function would make it quick. For the anisotropic functions
$\beta_\pm$ we have,

\begin{equation}
\rm \Delta \beta_\pm(\tau)=\pm \frac{2\kappa_\pm}{3\sqrt{{\kappa_\Omega }^2}}\,
arctanh\, \left(2\sqrt{\frac{3\Lambda}{{\kappa_\Omega }^2}}\left(
 -\frac{b_0}{48\Lambda}+e^{\sqrt{3\Lambda}\Delta \tau}\right) \right)
\end{equation}
and for the  $\phi $ function
\begin{equation}
\rm \phi(\tau)=\left\{
\begin{tabular}{ll}
$\rm \left[-
(m+2)\sqrt{\frac{\eta}{2\omega}}\frac{8}{\sqrt{{\kappa_\Omega }^2}}arctanh\,
\left(2\sqrt{\frac{3\Lambda}{{\kappa_\Omega }^2}}\left(
 -\frac{b_0}{48\Lambda}+e^{\sqrt{3\Lambda}\Delta \tau}\right) \right)\right]^{\frac{2}{m+2}}$,
& $f(\phi)=\omega
\phi^m, \quad m\not= -2$\\
$\rm
Exp\left\{-\sqrt{\frac{2\eta}{\omega}}\frac{8}{\sqrt{{\kappa_\Omega }^2}}arctanh\,
\left(2\sqrt{\frac{3\Lambda}{{\kappa_\Omega }^2}}\left(
 -\frac{b_0}{48\Lambda}+e^{\sqrt{3\Lambda}\Delta \tau}\right) \right)\right\}$, &
$f(\phi)=\omega
\phi^{-2}, \quad m= -2$\\
$\rm \frac{2}{m}\,Ln\left[-m\sqrt{\frac{ \eta}{2\omega}}
\frac{8}{\sqrt{{\kappa_\Omega }^2}}arctanh\,
\left(2\sqrt{\frac{3\Lambda}{{\kappa_\Omega }^2}}\left(
 -\frac{b_0}{48\Lambda}+e^{\sqrt{3\Lambda}\Delta \tau}\right) \right)\right]$, & $f(\phi)=\omega e^{m\phi}, \quad
m\not=0$\\
$\rm -\sqrt{2 \eta} \frac{8}{\sqrt{{\kappa_\Omega }^2}}arctanh\,
\left(2\sqrt{\frac{3\Lambda}{{\kappa_\Omega }^2}}\left(
 -\frac{b_0}{48\Lambda}+e^{\sqrt{3\Lambda}\Delta \tau}\right) \right)$& $f(\phi)=\omega, \quad m=0$
\end{tabular}
\right. \label{general-solution-phi2}
\end{equation}

\item{} {$\Lambda\not=0$ and $\gamma=1$}

Equation (\ref{omega-sol}) becomes
\begin{equation}\rm
\Delta \tau =\int \frac{12}{\sqrt{b_4e^{- 6\Omega} +48\Lambda}}
d\Omega \label{int2}
\end{equation}
where $\rm b_4={\kappa_\Omega }^2+48\mu_1$. In this case also we
have two possible solutions depending on the value of the
cosmological constant
\begin{itemize}
\item{} $\rm \Lambda>0$ .\\
The solution become
\begin{equation}\rm
\Delta
\tau=\frac{1}{\sqrt{3\Lambda}}\,arcsinh\,\left(4\sqrt{\frac{3\Lambda}{b_4}}e^{3\Omega}
\right)\label{sint2}
\end{equation}
so, the function $\Omega$ is
\begin{equation}
\rm
\Omega=\frac{1}{3}Ln\left|\frac{1}{4}\sqrt{\frac{b_4}{3\Lambda}}\,sinh\,
\left(\sqrt{3\Lambda}\Delta\tau\right)\right|
\end{equation}
and
\begin{eqnarray}
\rm e^{-3\Omega}&=&\rm 4\sqrt{\frac{3\Lambda}{b_4}}\, csch \, \left(\sqrt{3\Lambda}\Delta\tau \right)\nonumber\\
   \rm \int e^{-3\Omega}d\tau &=& \rm \frac{4}{\sqrt{b_4}}\, Ln
\left[tanh\,\left(\sqrt{3\Lambda}\Delta\tau\right)\right].\label{inv-1}
\end{eqnarray}
for this case the isotropization is possible, the corresponding
volume function would make it quick. For the anisotropic functions
$\beta_\pm$ we have
\begin{equation}
\rm \Delta\beta _{\pm}=\pm \frac{\kappa_{\pm} }{3\sqrt{b_4}}Ln
\left[tanh\,\left(\sqrt{3\Lambda}\Delta\tau\right)\right]
\end{equation}
and for the $\phi $ field
\begin{equation}
\rm \phi(\tau)=\left\{
\begin{tabular}{ll}
$\rm \left[- (m+2)\sqrt{\frac{\eta}{2\omega}}\frac{4}{\sqrt{b_4}}Ln
\left[tanh\,\left(\sqrt{3\Lambda}\Delta\tau\right)\right]\right]^{\frac{2}{m+2}}$,
& $f(\phi)=\omega
\phi^m, \quad m\not= -2$\\
$\rm Exp\left\{-\sqrt{\frac{2\eta}{\omega}}\frac{4}{\sqrt{b_4}}Ln
\left[tanh\,\left(\sqrt{3\Lambda}\Delta\tau\right)\right]\right\}$,
& $f(\phi)=\omega
\phi^{-2}, \quad m= -2$\\
$\rm \frac{2}{m}\,Ln\left[-m\sqrt{\frac{\omega \eta}{2\omega}}
\frac{4}{\sqrt{b_4}}Ln
\left[tanh\,\left(\sqrt{3\Lambda}\Delta\tau\right)\right]\right]$, &
$f(\phi)=\omega e^{m\phi}, \quad
m\not=0$\\
$\rm -\sqrt{2 \eta} \frac{4}{\sqrt{b_4}}Ln
\left[tanh\,\left(\sqrt{3\Lambda}\Delta\tau\right)\right] $&
$f(\phi)=\omega, \quad m=0$
\end{tabular}
\right. \label{general-solution-phi3}
\end{equation}

\item $\rm \Lambda<0$ .\\
The corresponding solutions are
\begin{equation}\rm
\Delta \tau=-\frac{1}{\sqrt{3|\Lambda|}}\,
arccos\,\left(4\sqrt{\frac{3|\Lambda|}{b_4}}\,e^{3\Omega}
\right)\label{sint22}
\end{equation}
the function  $\Omega$
\begin{equation}\rm
\Omega=\frac{1}{3} Ln \left|\frac{1}{4}
\sqrt{\frac{b_4}{3|\Lambda|}} cos\,
\left(\sqrt{3|\Lambda|}\Delta\tau\right)\right|
\end{equation}
as the volume has an oscillatory behavior, the isotropization do not
yield for this case, and for completeness we calculate
\begin{eqnarray}
\rm e^{-3\Omega}&=&\rm 4\sqrt{\frac{3|\Lambda|}{b_4}}\sec\, \left(\sqrt{3|\Lambda|}\Delta\tau\right)\nonumber\\
\rm \int e^{-3\Omega} d\tau&=&\rm \frac{4}{\sqrt{b_4}}
Ln\left|\sec\left(\sqrt{3|\Lambda|}\,\Delta\tau\right)+\tan\left(\sqrt{3|\Lambda|}\,\Delta\tau\right)\right|\nonumber \\
&=&\rm \frac{4}{\sqrt{b_4}}Ln\left| \frac{\cos
\left(\frac{\sqrt{3|\Lambda|}}{2}\Delta\tau\right)+ \sin
\left(\frac{\sqrt{3|\Lambda|}}{2}\Delta\tau\right)} {\cos
\left(\frac{\sqrt{3|\Lambda|}}{2}\Delta\tau\right)- \sin
\left(\frac{\sqrt{3|\Lambda|}}{2}\right)}\right|,
\end{eqnarray}
the anisotropic functions $\beta_\pm$,
\begin{equation}\rm
\Delta\beta _{\pm}=\mp \frac{\kappa_{\pm}
}{3\sqrt{b_4}}Ln\left|\sec\left(\sqrt{3|\Lambda|}\,\Delta\tau\right)+\tan\left(\sqrt{3|\Lambda|}\,\Delta\tau\right)\right|
\end{equation}
and the $\phi $ field
\begin{equation}
\rm \phi(\tau)=\left\{
\begin{tabular}{ll}
$\rm \left[ (m+2)\sqrt{\frac{\eta}{2\omega}}\frac{4}{\sqrt{b_4}}
Ln\left|\sec\left(\sqrt{3|\Lambda|}\,\Delta\tau\right)+\tan\left(\sqrt{3|\Lambda|}\,\Delta\tau\right)\right|\right]^{\frac{2}{m+2}}$,
& $f(\phi)=\omega
\phi^m, \quad m\not= -2$\\
$\rm
Exp\left\{\sqrt{\frac{2\eta}{\omega}}\frac{4}{\sqrt{b_4}}Ln\left|\sec\left(\sqrt{3|\Lambda|}\,\Delta\tau\right)+\tan\left(\sqrt{3|\Lambda|}\,\Delta\tau\right)\right|\right\}$,
& $f(\phi)=\omega
\phi^{-2}, \quad m= -2$\\
$\rm \frac{2}{m}\,Ln\left[m\sqrt{\frac{ \eta}{2\omega}}
\frac{4}{\sqrt{b_4}}Ln\left|\sec\left(\sqrt{3|\Lambda|}\,\Delta\tau\right)+\tan\left(\sqrt{3|\Lambda|}\,\Delta\tau\right)\right|\right]$,
& $f(\phi)=\omega e^{m\phi}, \quad
m\not=0$\\
$\rm \sqrt{2 \eta}
\frac{4}{\sqrt{b_4}}Ln\left|\sec\left(\sqrt{3|\Lambda|}\,\Delta\tau\right)+\tan\left(\sqrt{3|\Lambda|}\,\Delta\tau\right)\right|$&
$f(\phi)=\omega, \quad m=0$
\end{tabular}
\right. \label{general-solution-phi4}
\end{equation}

\end{itemize}
\end{enumerate}


\section{Final remarks}
In this work we present the study of the classical cosmological
anisotropic Bianchi type I in the K-essence formalism. In previous
work made by Chimento
 and co-research   \cite{chimento}, they present the possible isotropization of this
model.  Our goal in this work is that we obtain the corresponding
classical solutions for a barotropic perfect fluid and cosmological
term $\Lambda$ who mimetic the scalar field in equation (\ref{kk}).
In the case of $\Lambda=0$ and $\gamma\not=1$ we obtain the
solutions in closed form. With this solutions we can validate our
qualitative analysis on isotropization of the cosmological model,
implying that this become when the volume is large in the
corresponding {\it time} evolution. So, only one solutions do not
present the large volume, when $\Lambda<0$ in stiff matter era in
the ordinary matter content. We include a qualitative analysis to
Friedmann equation when it is written as an equation for the volume
that is equivalent to the equation of motion of a particle under a
potential and we conclude the same about the isotropization of this
anisotropic model, considering the stiff matter and radiation cases.
In the quantum analysis for this model, considering the scalar
field, the solutions are similar that those found in the Bianchi
type IX cosmological model \cite{bianchi-ix}, you can see the
equation (29). For quantum radiation case, the resulting
Wheeler-DeWitt equation appears as fractionary differential
equation, and the results will be reported elsewhere.

\acknowledgments{ \noindent This work was partially supported by
CONACYT  167335, 179881 grants. PROMEP grants UGTO-CA-3 and
UAM-I-43. This work is part of the collaboration within the
Instituto Avanzado de Cosmolog\'{\i}a and Red PROMEP: Gravitation
and Mathematical Physics under project {\it Quantum aspects of
gravity in cosmological models, phenomenology and geometry of
space-time}. Many calculations where done by Symbolic Program REDUCE
3.8.}

\end{document}